\documentclass[aps,prl,superscriptaddress,amsmath,amssymb,amsfonts,twocolumn,showpacs,10pt]{revtex4-1}
\usepackage{amssymb,verbatim}
\usepackage{amsmath}
\usepackage{graphicx}
\usepackage{color}
\usepackage[dvipsnames]{xcolor}
\usepackage[colorlinks=true,citecolor=blue]{hyperref}
\usepackage{multirow}
\usepackage{fourier,charter}
\graphicspath{{figures/}}
\usepackage{cancel}
\usepackage[normalem]{ulem}

\newcommand{\orcid}[1]{\href{https://orcid.org/#1} 
  {\includegraphics[width=10pt]{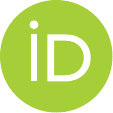}}}

\def\sech{\mathop{\rm sech}\nolimits}

\def\be{\begin{equation}}
\def\ee{\end{equation}}

\def\HW[#1]{{\small\normalfont[#1]}}
\let\~=\tilde

\newcommand{\dd}{{\rm d}}

\def\bse{\begin{subequations}}
\def\ese{\end{subequations}}

\newcommand{\ceff}{c^{{\rm eff}}}
\newcommand{\veff}{v^{{\rm eff}}}

\advance\textwidth 4mm
\advance\hoffset -2mm
\advance\textheight 4mm
\advance\voffset -2mm

\begin{document}

\title{Kinetic equations for a two-dimensional soliton gas}
\author{Gino Biondini\orcid{0000-0003-3835-1343}}
\affiliation{Department of Mathematics, State University of New York, Buffalo, NY, United States of America}
\author{Thibault Bonnemain\orcid{0000-0003-0969-2413}}
\affiliation{Laboratoire de Physique Théorique et Modélisation, CNRS UMR 8089,
CY Cergy Paris Université, 95302 Cergy-Pontoise Cedex, France}
\author{Benjamin Doyon\orcid{0000-0002-5258-5544}}
\affiliation{Department of Mathematics, King's College, London, United Kingdom}
\author{Gennady  El\orcid{0000-0003-1962-5388}}
\author{Giacomo Roberti\orcid{0000-0001-8233-9531}}
\affiliation{School of Engineering, Physics and Mathematics, Northumbria University, Newcastle upon Tyne, United Kingdom}

\date{\small\today}

\begin{abstract}
We formulate a general system of kinetic equations for a non-stationary two-dimensional gas of elastically interacting line solitons and apply it to the description of a soliton gas governed by the Kadomtsev–Petviashvili II (KPII) equation. We then verify the predictions of the kinetic theory in two analytically tractable problems: the oblique interaction of a KPII line soliton with a one-dimensional soliton condensate of the Korteweg–de Vries equation, and the interaction of a trial KPII soliton with a monochromatic KPII soliton gas. In both cases, we compare the analytical results with direct numerical simulations obtained by constructing two-dimensional soliton gases via exact KPII $N$-soliton solutions for large $N$, using appropriately chosen random distributions of soliton parameters. The comparison demonstrates excellent agreement, thereby providing strong validation of the proposed kinetic theory of  2D non-equilibrium soliton gases.
\end{abstract}

\maketitle


The concept of a soliton gas (SG)—an infinite, random ensemble of interacting solitons—has gained attention as a powerful framework for analyzing random nonlinear wave fields through the lenses of statistical physics and hydrodynamics of integrable systems. Initiated by Zakharov in \cite{zakharov1971kinetic}, SG theory has grown into  an area of intense mathematical and physical research, providing deep insights into long-standing problems of nonlinear physics such as spontaneous modulational instability, wave-meanflow interaction and rogue wave formation \cite{gelash2018strongly,gelash2019bound,Agafontsev_2015,PhysRevLett.132.207201,congy2026exactly}. Experimental and observational evidence of soliton gases has  been reported in a range of physical settings, including surface water waves \cite{Costa:14, redor2019experimental, Suret2020Nonlinear}, photorefractive crystals \cite{marcucci_topological_2019-1}, and superfluid systems \cite{mossman_observation_2023}.  For the state-of-the-art theoretical and experimental developments on SGs, please see the recent reviews \cite{el2021soliton,suret_soliton_2024}.

At the heart of SG theory lies the kinetic equation for the evolution of the density of states (DOS)---the joint distribution of solitons with respect to their amplitudes and positions. This equation has been derived for integrable disperive hydrodynamic models such as the Korteweg–de Vries (KdV) and the nonlinear Schrödinger (NLS) equations \cite{el2003thermodynamic, el2020spectral} using the spectral framework of the inverse scattering transform (IST), where individual solitons correspond to discrete eigenvalues of the associated linear Lax operator. These systematic spectral derivations, in turn, have motivated a direct phenomenological formulation of the SG kinetic equations based on postulating the {\it collision rate ansatz} \cite{el2005kinetic,doyon2018soliton,doyon2019generalized,congy2021soliton,bonnemain2022generalized,bonnemain2025soliton}. In this approach, solitons are treated as quasiparticles undergoing pairwise, short-range interactions accompanied by phase/position shifts. The cumulative effect of these interactions leads to an emergent hydrodynamic or kinetic description on the macrocopic, Eulerian scale.

A related strand of research  is  {\it generalised hydrodynamics} (GHD), where equations analogous to the SG kinetic equation have been derived to describe the emergent hydrodynamic behaviour of integrable quantum and classical many-body systems \cite{castro2016emergent, bertini2016transport, doyon2020lecture, doyon_generalized_2025}. As it turns out, the GHD framework applies equally well to SGs in integrable dispersive hydrodynamic systems. 
The GHD of SGs for the KdV and Boussinesq equations was recently
developed in \cite{bonnemain2022generalized} and \cite{bonnemain2025soliton} respectively, and that of NLS (as a semiclassical limit of the Lieb-Liniger model) in \cite{koch2022generalized}.
Beyond providing a kinetic/hydrodynamic description, GHD also enables  formulation of SG thermodynamics—including free energy, entropy, and temperature—through the fundamental concepts of the Thermodynamic Bethe Ansatz and the Generalized Gibbs Ensemble. 

So far, both the spectral SG theory and GHD have been mostly restricted to one-dimensional (1D) systems.  However, while providing a deep insight into many nonlinear wave phenomena, the 1D framework is often too restrictive.  Examples of inherently multidimensional soliton-bearing nonlinear wave phenomena include complex line soliton interaction patterns  ubiquitously observed in water tank experiments  and in nearshore waters \cite{chakravarty_kp_2014, ablowitz_nonlinear_2012}, particularly in crossing sea states \cite{pelinovsky_non-gaussian_2016},  the generation of 2D and 3D solitary waves and  undular bores (i.e., dispersive shock waves, DSWs) in the stratified ocean \cite{vlasenko_three-dimensional_2009, yuan_propagation_2018}, in spatial nonlinear optics \cite{wan_dispersive_2007} and superfluids \cite{el_oblique_2006, hoefer_dispersive_2006, hoefer_oblique_2017}. 
An experimental set up to generate and observe (random) 2D SGs in shallow water was also recently developed \cite{leduque2024space}; the observation of a 2D SG formation in a photon fluid was reported in \cite{dieli2024observation}. These physical examples provide a strong motivation for the development of the theory of 2D SGs, which is the subject of the present Letter.

Extending SG and GHD theories to the spatially two-dimensional setting is challenging because of the increased complexity of multidimensional nonlinear wave interactions.
The first step in this direction was undertaken by the present authors in \cite{bonnemain2025two}, where the theory of a 2D stationary SG was constructed within the framework of the time-independent reduction of the Kadomtsev–Petviashvili  (KP) equation \cite{kadomtsev1970stability}. The KP equation represents the simplest fundamental generalization of the KdV equation to weakly two-dimensional nonlinear dispersive waves, 
and is a universal model for weakly two-dimensional nonlinear dispersive waves in the long wavelength regime \cite{ablowitz1981solitons}. It arises in two canonical forms: KPI and KPII.
Relevant to this Letter is the KPII equation, which supports stable line solitons and models two-dimensional nonlinear wave propagation across a wide range of physical settings—from shallow water waves and plasma physics \cite{kodama_solitons_2013, Ruderman_2020} to ferromagnetism and Bose–Einstein condensates \cite{turitsyn_stability_1985, PhysRevA.67.023604, kamchatnov_stabilization_2008, hoefer_dark_2012}. 

The stationary 2D SG construction in \cite{bonnemain2025two} is based on the equivalence between the time-independent (2+0)-D KP equation and the classical (1+1)-D integrable good Boussinesq  equation. This correspondence effectively reduces the problem to a 1D setting, allowing one to invoke the existing 1D theory for bidirectional SGs \cite{bonnemain2025soliton,congy2021soliton}, with the transverse spatial variable $y$ assuming the role of an evolution (time-like) variable.

\begin{figure}[t!]
\centerline{\includegraphics[trim=0 0 5 5,clip,width=1.025\linewidth]{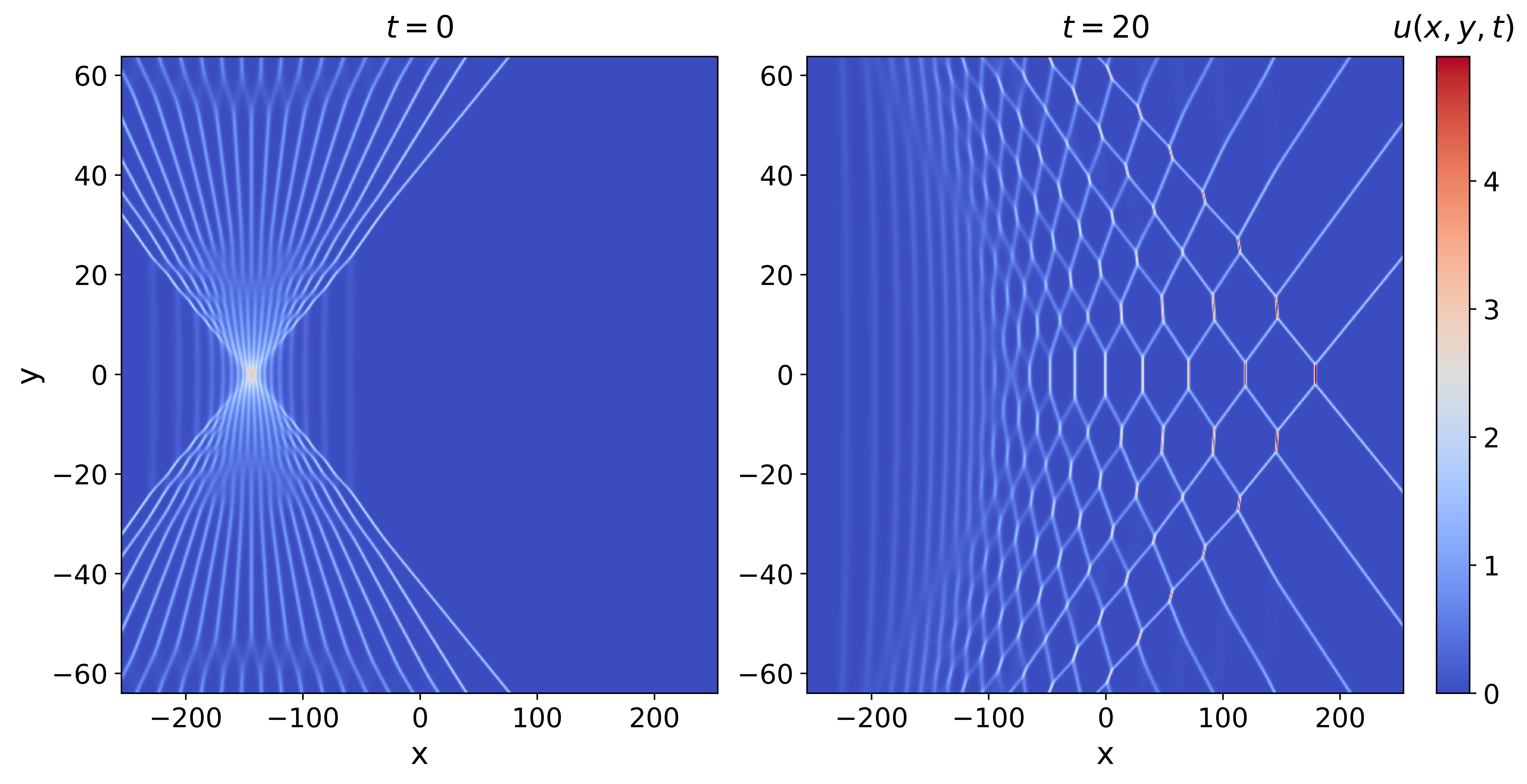}}
\vglue\smallskipamount
\centerline{\includegraphics[trim=0 0 5 5,clip,width=1.025\linewidth]{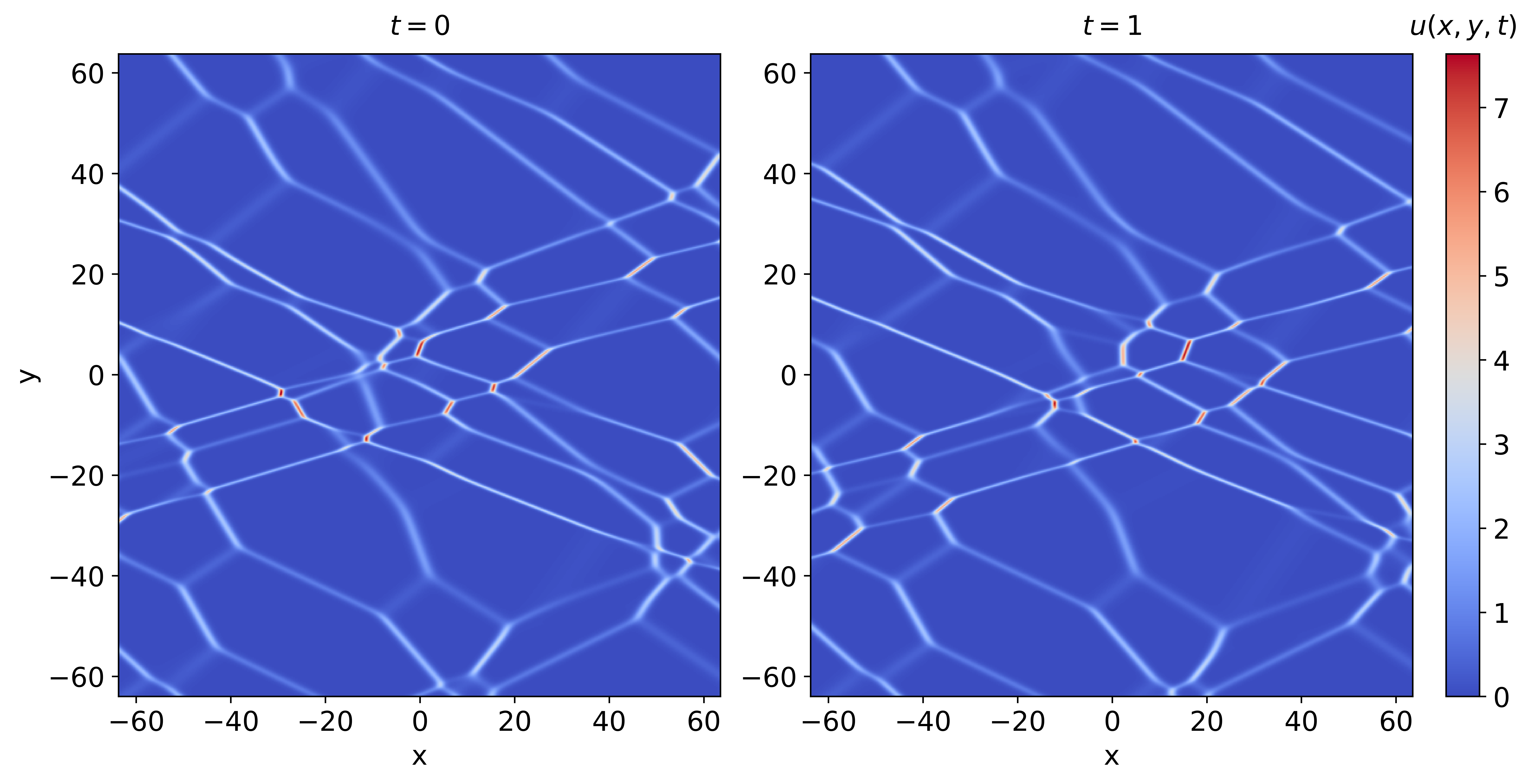}}
\vglue-\smallskipamount
\caption{Deterministic versus stochastic non-stationary 2D SG realized via  exact KPII $N$-soliton solutions. 
Top row: snapshots at $t=0$ (left) and $t=20$ (right) of a deterministic SG with $N=40$ and fixed initial phase, amplitude and slope parameters.
Bottom row: snapshots at $t=0$ (left) and $t=1$ (right) of a realization of a stochastic SG  with $N=30$ and random initial phase, amplitude and slope parameters.
See \cite{supplement} for  details.}
\label{f:1}
\kern-\smallskipamount
\end{figure}

In this Letter we present the first analytical description of the emergent hydrodynamics of a  \textit{non-stationary} (2+1)-dimensional SG.  We first formulate the general system of kinetic equations for a 2D integrable {\it gas of line solitons} using GHD principles.  Then we apply this system to the particular case of the dense gas of line solitons of the KPII equation for which the basic soliton interaction parameter---the two-soliton scattering position shift---was obtained in \cite{biondini2007line}. 

We  use the KPII SG kinetic equations to consider two  analytically tractable benchmark problems: (i)~oblique interaction of a line soliton with 1D coherent structures: rarefaction waves and DSWs generated by box-type initial conditions and interpreted as modulated critically dense SGs (soliton condensates) of the KdV equation  \cite{congy2023dispersive}; (ii)~interaction of a ``trial'' (or test) soliton with a ``monochromatic'' SG with a DOS given by a 2D delta-function.  In both cases the kinetic theory gives exact predictions for the 2D evolution of the KP soliton interacting with SG.  We verify these predictions numerically  by building the corresponding KP SG  via exact $N$-soliton solutions \cite{biondini2006soliton,freeman1983soliton,hirota2004direct,kodama2017kp} 
with large $N$ and appropriately chosen random distributions for the soliton parameters. Two examples of numerical realizations of 2D KP SGs are shown in Fig.~\ref{f:1}.

%

\medskip

\paragraph{Dynamics and hydrodynamics of line solitons.}
A crucial characteristic of a large class of higher-dimensional integrable systems is the presence of \textit{co-dimension-1} solitons. Recall that, in one spatial dimension, a soliton is a spatially localized wave that propagates linearly in time up to interactions with other solitons.  Viewed from ``large scales'', this is a 0-dimensional, point-like object. Likewise, in $d$-dimensional spaces, a co-dimension 1 soliton is a wave whose support is concentrated on a region that extends only in $d-1$ dimensions, and which, up to interactions, propagate in time by a linear displacements in the direction perpendicular to this region. In dimensions $d>1$ such extended objects may have complicated geometry, but the simplest are ``planar objects,'' waves supported on $d-1$-dimensional hyper-planes in $d$ dimensions. For the KP equation, with $d=2$, these are line solitons.  Since this is the model of interest here, and for simplicity, we next concentrate on line solitons in $d=2$, but the extension to $d>2$ is immediate.

The main question of interest here is: 
What is the effect of interactions when many line solitons are present? In~particular, as extended solitons of different slopes always cross in space, what happens at the crossings? This question may be understood from the integrability structure. Recall that, for integrable systems, there is a hierarchy of ``higher times'' that form commuting flows. These include the usual time which we call $t$ here, as well as other times $t_1,\,t_2,\ldots$, with $x = t_1$. \textit{A line soliton in the $xy$-plane can be seen as the trajectory of a usual point-like soliton under one such higher time, say $t_2$, which is equated to the $y$ coordinate}. The flow under time $t = t_3$ then describes the time evolution of these line solitons. Now, in one dimension, it is well known that integrable time evolution gives rise to elastic, factorised scattering: solitons keep their shapes and velocities after interactions (elastic scattering), being only affected by spatial shifts of their trajectories that, in a many-soliton interaction, add up as if they had interacted 2 solitons at a time (factorised scattering). This is true for any time in the hierarchy. This is how line solitons interact in two spatial dimensions: besides having complicated wave forms at crossings in the plane, they otherwise keep their slopes~$c$ in the $xy$-plane and move with time $t$ along the $x$ direction at fixed velocity~$v$.   A parametrisation of the velocity that useful in view of the application to the KPII equation is that using the ``amplitude parameter''~$a$,
\vspace*{-0.8ex}
\be\label{va}
v = 4a^2 + c^2\,.
\ee
At crossings with another line soliton $(a',c')$, the line soliton $(a,c)$  is affected by a finite spatial shift $\varphi(a,c;a',c')$ in the $x$ direction, which has a factorised form at multiple crossings, in accordance to factorised scattering theory.

A crucial point is that, although there are two ``times'' at play ($y=t_2$ and $t= t_3$), by consistency of the commuting flows there is a single shift function $\varphi(a,c;a',c')$ that describes the shift occurring for both time evolutions.  It~is simple to see why: imagine two line solitons in the $xy$-plane at time 0, crossing at one point $(x_0,y_0)$, say with $y_0<0$. Take a section, say $y=0$, and assume that, on that section, the line solitons' $x$-coordinates approach each other in time and cross at some time $t_0>0$. Then the spatial crossing point $(x_{t},y_{t})$ has moved to the other side of the section, $y_{t}>0$ for $t>t_0$. As line solitons keep their slopes and velocities, shifts that occur at spatial crossings do not vary in time. Therefore, by pure geometric reasoning, the passage of line-soliton crossing from $y<0$ to $y>0$ in time, tells us exactly that the scattering shift due to time evolution must be the same as that from the spatial crossing itself.

We use this simple picture of line solitons in order to write down a theory for a {\em dense line-soliton gas}. In any configuration with, say, $N$ line solitons, in the asymptotic region of the $xy$-plane the line solitons do not interact and are easily identifiable, describing straight lines with slopes $c_1,c_2,\ldots,c_N$. Linearly extending the $y\to-\infty$ region of line soliton $i$ to the section $y=0$, its $x$ coordinates gives its ``incoming impact parameter'' $x_i^-(t)$. This moves linearly in time, $x_i^-(t) = x_i^-(0) + v_i t$. In finite regions of the $xy$-plane and at finite times, however, where line solitons interact, they are in general not easily identifiable, because interactions completely modify the wave form. Nevertheless, following the theory of soliton gases in one dimension \cite{suret_soliton_2024}, we conjecture that the approximate shape of line solitons in the $xy$-plane determined by the collision rate ansatz for the $y=t_2$ evolution, gives a correct description of the wave field on large scales, for a large class of impact parameters. Therefore, at large scales, {\em we may describe the wave form statistics by giving the density of line solitons}. For this purpose, we define
\vspace*{-0.4ex}
\begin{equation}\label{dos}
    \rho_{a,c}(x,y,t)\dd a\dd c\dd x
\end{equation}
as the number of line solitons with asymptotic slope in $[c,c+\dd c]$, amplitude parameter in $[a,a+\dd a]$ and $x$-intercept on the section $y$ in $[x,x+\dd x]$, at time $t$. 
The quantity $\rho_{a,c}(x,y,t)$ is the 2D counterpart of the {\em density of states} in the spectral theory of 1D SGs \cite{el2003thermodynamic,  el2005kinetic, el2020spectral,  el2021soliton, suret_soliton_2024}, where it satisfies the continuity equation consistent with isospectrality of integrable evolution within the IST framework.  Then, the theory of 1D soliton gases immediately implies the following {\em kinetic equations} for the commuting flows
\bse
\label{conty}
\begin{gather}
    \partial_y \rho_{a,c} + \partial_x\left(\ceff(a,c)\rho_{a,c}\right) = 0,
    \\
    \partial_t \rho_{a,c} + \partial_x\left(\veff(a,c)\rho_{a,c}\right) = 0 \ ,
\end{gather}
\ese
along with the equations of state following from the 1D collision rate ans\"atze for $xt$- and $xy$- evolutions \cite{el2005kinetic,doyon2018soliton,doyon2019generalized,congy2021soliton,bonnemain2022generalized,bonnemain2025soliton}:
\begin{equation}\label{ceff}
\begin{aligned}
    \ceff(a,c) &= c + \\
    & \int_{A\times C} \dd a'\dd c' \rho_{a',c'}
      \varphi(a,c;a',c') \left[\ceff(a',c')-\ceff(a,c)\right] \ .
\end{aligned}
\end{equation}
{\em and the same equation with $\ceff(a,c)$ replaced by $\veff(a,c)$ and $c$ by $v=4a^2+c^2$}. 
Here $A\times C$ is the DOS' support.
The self-consistency of Eqs.~\eqref{conty} requires 
\begin{equation}\label{consist}
\partial_t \left(\ceff(a,c)\rho_{a,c}\right) = \partial_y \left(\veff(a,c)\rho_{a,c}\right)\, .
\end{equation} 
Indeed, in the End Matter we show that equation \eqref{consist} holds  for the general gas of line solitons. 
Also in the End Matter we give an alternative derivation of Eqs.~\eqref{conty} and~\eqref{consist} based on a thermodynamic limit of modulation equations for multiphase solutions of  integrable 2D dispersive hydrodynamics.
We also note that the above equations reduce to those in \cite{bonnemain2025two} in the case of a stationary KPII SG.
Next we apply this general theory to the non-stationary SG of the KPII equation.


%
The KP equation is the nonlinear PDE \cite{kadomtsev1970stability}
\vspace*{-0.4ex}
\be
(u_t + 6 u u_x + u_{xxx})_x + \sigma u_{yy} = 0\,,
\label{e:KP}
\ee
where $u=u(x,y,t)$ is a real-valued field and 
subscripts $x,y,t$ denote partial differentiation.
The constant $\sigma=\pm1$ distinguishes between the  KPI and KPII variants of the equation, respectively.
The numerical coefficients in front of the various terms can be changed by simple rescalings of the dependent and independent coordinates.
The KP equation is the prototypical completely integrable system in two spatial dimensions and its initial value problem with localized fields is solvable via the IST \cite{ablowitz1991solitons}.
On the other hand, the treatment of solitons in the IST for the KP equation is still in some respects a subject of current research \cite{wu2021direct}.
While both the KPI and KPII equations admit line soliton solutions, those of the KPI equation are unstable 
to transverse perturbations \cite{pelinovsky1993self},
while those of KPII are linearly stable.  Moreover, the KPII equation admits a large  family of multi-soliton solutions displaying soliton resonance and web structure \cite{biondini2007line,biondini2006soliton,biondini2003family,kodama2004young}, 
as well as a surprising and deep connection with algebraic combinatorics \cite{kodama2017kp}. 

Line solitons of the KPII equation on a zero background have the form
\vspace*{-0.4ex}
\be
u(x,y,t) = 2a^2\sech^2[a(x - c y - v t - x_0]\,,
\label{e:KPsoliton}
\ee
where $x_0 = x^-(0)$, and the soliton dispersion relation \eqref{va} holds.
In a two-soliton solution, if the resonance condition
\be\label{eq:rescond}
-(a_1+a_2) \leq (c_2 - c_1)/(2\sqrt3) \leq a_2- a_1,
\ee
holds \cite{biondini2007line},
the interaction solitons $(a_1,c_1)$ and $(a_2,c_2)$ is resonant, and the solitons merge to produce four Y-shape Miles soliton resonances \cite{miles1977diffraction,biondini2003family}. 
Otherwise, the interaction is non-resonant, and results in a spatial shift \
\vspace*{-0.4ex}
\begin{equation}
    \label{e:phaseshifts}
    \varphi(a,c;a',c') = 
\frac{1}{2 a}
    \log \left|\frac{(c-c')^2 - 12(a-a')^2}
        {(c-c')^2 - 12(a+a')^2}\right|\,.
\end{equation}
A positive (negative) value implies soliton $(a,c)$ shifts to the left (right) if it meets $(a',c')$ to its right, and vice versa. This phenomenon generalises pairwise to $N$-soliton solutions \cite{kodama2004young,biondini2007line}.
Thus, Eqs.~\eqref{conty}--\eqref{consist}, describing a gas of line solitons, are expected to be the large-scale kinetic equations for the KPII equation that arise in non-resonant $N$-soliton solutions, where each pair of solitons satisfies the non-resonance condition given by the opposite of~\eqref{eq:rescond}.


To verify validity of the kinetic equations we consider two benchmark examples.
The first one is the oblique interaction of a trial line soliton with 1D rarefaction waves and DSWs generated by the box-type initial conditions and interpreted as modulated critically dense SGs---soliton condensates---of the KdV equation (note that the KdV equation is an exact one-dimensional reduction of KPII so the KdV solution also satisfies the KP) \cite{congy2023dispersive}.  Let the KdV condensate DOS $\rho(a,x,t)$ be supported on $a \in [0,1]$.   The effective slope of the KPII oblique trial soliton with parameters $a_{*}>1$ and $c_{*}<0$ interacting with the KdV condensate is then found from \eqref{ceff} to be 
\begin{equation}\label{ceff_test1}
    \ceff(a_{*},c_{*}) = \frac{c_{*}}{1 - \int_0^1\dd a'\, \rho(a',x,t)\log\left(\frac{c_{*}^2-12(a_{*}-a')^2}{c_{*}^2-12(a_{*}+a')^2}\right) }\; ,
\end{equation}
where  $\rho(a,x,t)$ is defined according to the region shown in Fig.~\ref{f:oblique}, see \cite{supplement} for details. Comparison of the analytical prediction \eqref{ceff_test1} with the effective slope of the trial soliton in the numerically realized interaction with KdV soliton condensate is shown in Fig.~\ref{f:oblique} and demonstrates excellent agreement. Details of the numerical simulations can be found in \cite{supplement}.

The second example we consider is the interaction of a  oblique trial soliton with a ``monochromatic'' SG whose DOS given by a 2D delta-function.  
The kinetic equations \eqref{conty}--\eqref{consist} admit polychromatic reductions, under the ansatz $\rho_{a,c}(x,y,t) = \sum_{j=1}^M w_j(x,y,t)\delta(a-a_j)\delta(c-c_j)$, see the End Matter.
The interaction of an oblique $(a_*, c_*)$-soliton with a monochromatic gas with density $w_1$ and slope $c_1$ corresponds to $M=2$ and $w_2(x,y,t)=0$. Then the effective slope  of the trial soliton with $c_{*} = c_2$, $a_{*} = a_2$ is given by 
\begin{equation}\label{ceff_test2}
\ceff (a_{*}, c_{*})= \frac{c_{*} + w_1 \varphi_{21} c_{*}}{1+w_1\varphi_{21}}, \quad \varphi_{21} = \varphi(a_{*}, c_{*}; a_1, c_1),
\end{equation}
and a similar expression for $\veff(a_*, c_*)$, see Eq.~\eqref{bichrom_eff}b in the End Matter. Comparisons of the analytically predicted effective slope \eqref{ceff_test2} and the effective velocity of the trial soliton with those observed in numerical simulations of the interaction with monochromatic SG are shown in
Fig.~\ref{f:monochromatic} and, again, demonstrate excellent agreement.

The above two examples provide a strong evidence of the validity of the kinetic equations \eqref{conty}, \eqref{consist}, \eqref{ceff}.

\begin{figure}[t!]
    \centering
    \includegraphics[clip,width=\linewidth]{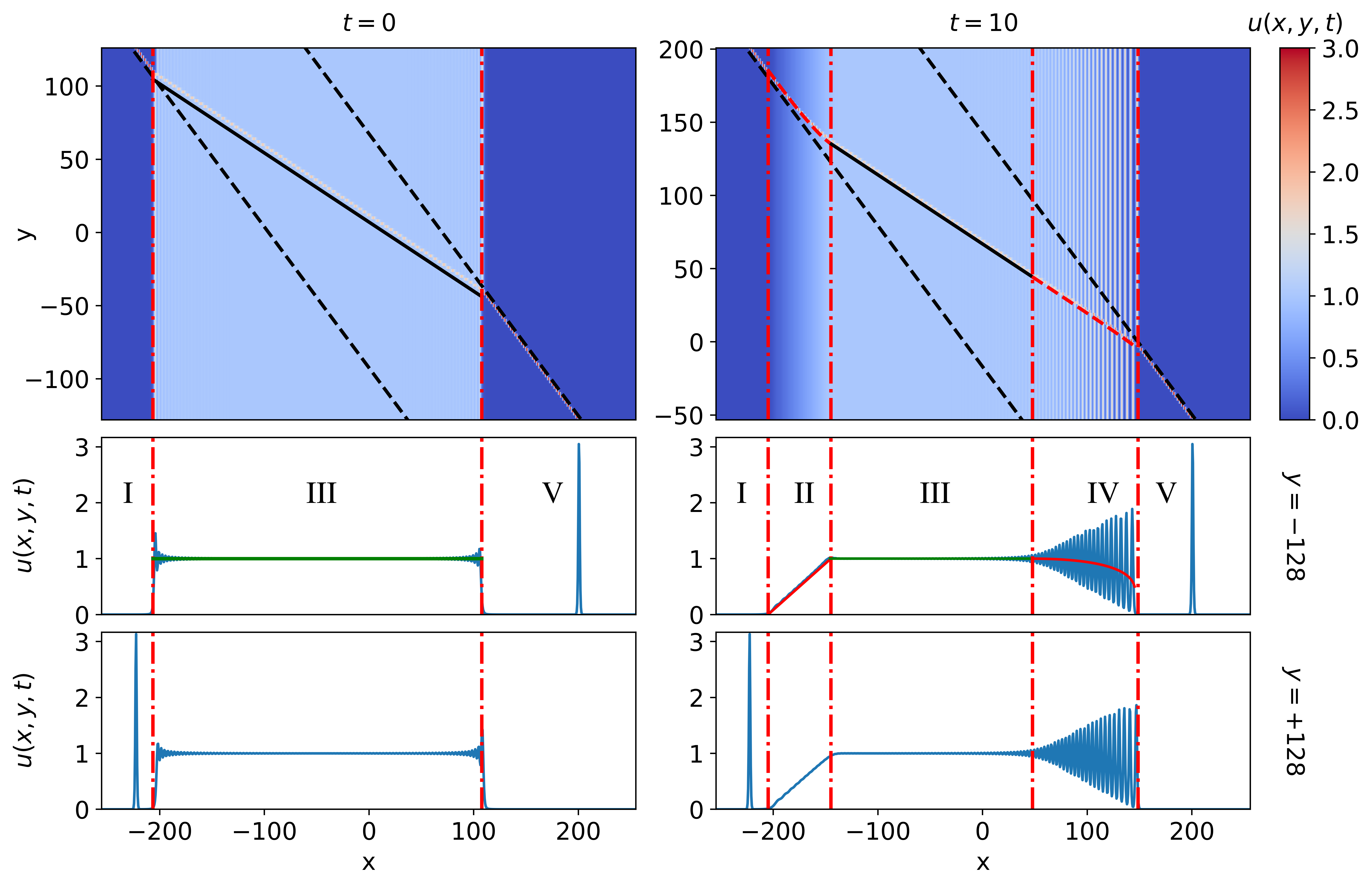}
    \caption{Oblique interaction of a trial KP soliton with localized KdV soliton condensate realized numerically via $N=100$-soliton KdV solution;  at $t=0$ (left) and $t=10$ (right).
    The black-dashed lines represent the slope of the trial line soliton in the absence of interaction. The analytically predicted effective slopes of the trial soliton resulting from the interaction with the soliton condensate are represented by black-solid lines in the genus zero condensate (III) region, and as red-dashed lines in the RW (II) and DSW (IV) regions described by the modulated genus zero and genus one condensates respectively.
    See \cite{supplement} for  details.}
    \label{f:oblique}
\bigskip
    \centering
    \includegraphics[trim=0 5 0 5,clip,width=1.025\linewidth]{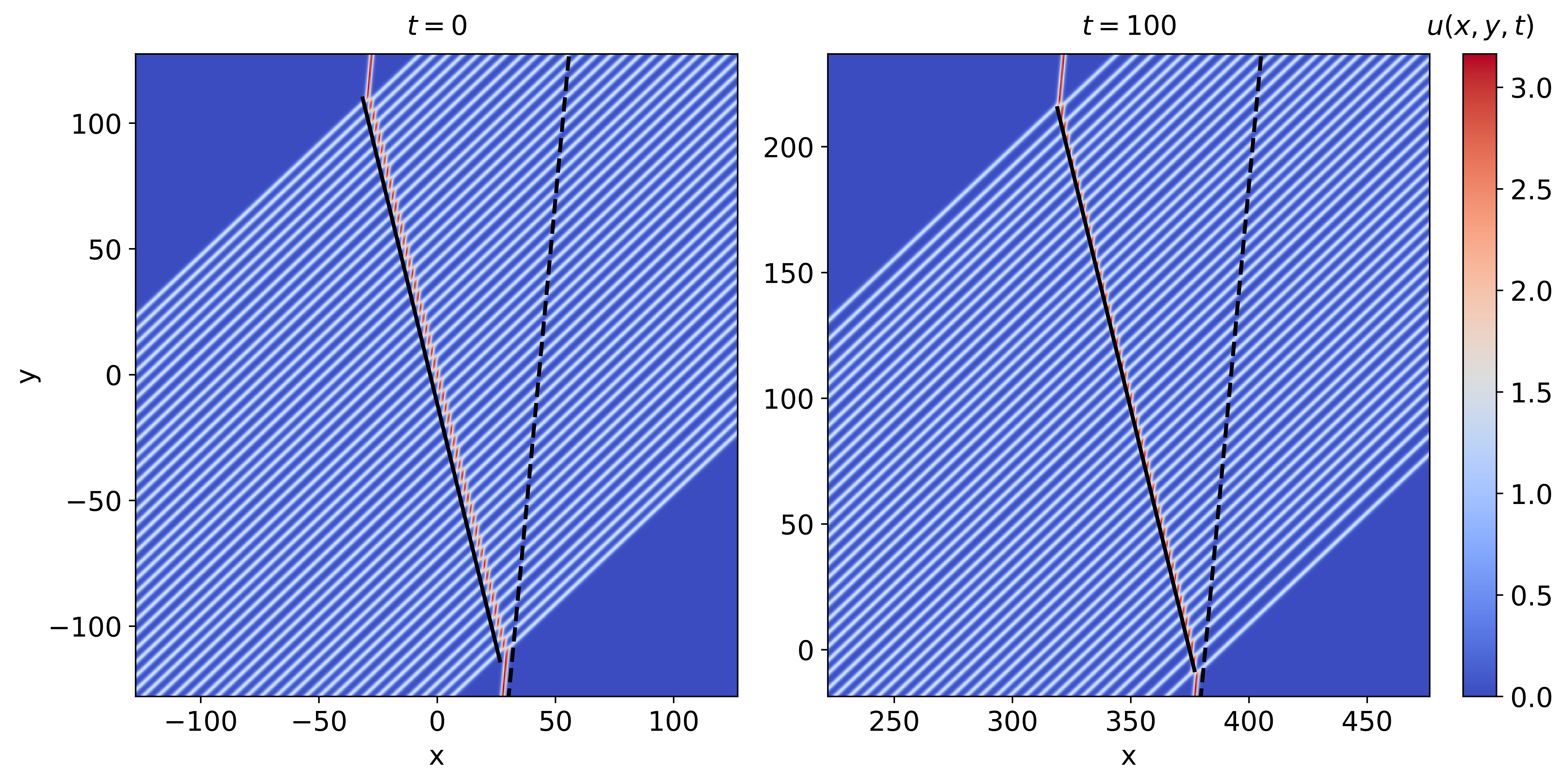}
    \caption{Refraction of a trial line soliton by a monochromatic KPII soliton gas realized via $N=50$-soliton solution; at $t=0$ (left) and $t=100$ (right). The black-dashed lines represent the slope of the noninteracting trial line soliton, and the black-solid lines represent the analytically predicted effective slope \eqref{ceff_test2} of the trial soliton (red line) resulting from the interaction with the SG. The position of the black-solid line in the right panel ($t=100$) is obtained by shifting its position at ($t=0$) by  $\Delta x = t v^{\rm eff}(a_{*},c_{*})$.
    See \cite{supplement} for  details.}
    \label{f:monochromatic}
\end{figure}

%
In conclusion, we formulated a general system of kinetic equations describing the emergent hydrodynamics of a non-stationary $(2{+}1)$-dimensional gas of line solitons. The construction rests on the identification of a line soliton in the $xy$-plane with the trajectory of an ordinary point-like soliton under a higher time $t_2=y$, together with the geometric observation that a single scattering shift 
governs the displacement under both the $t_2$ and the $t_3=t$ flows. 
In the case of the KPII equation, we benchmarked the theory against direct numerical simulations built from exact $N$-soliton solutions~\cite{biondini2006soliton,freeman1983soliton,hirota2004direct,kodama2017kp} with large~$N$ and random soliton parameters, considering both (i)~the oblique interaction of a trial line soliton with a KdV soliton condensate realizing a rarefaction wave or DSW~\cite{congy2023dispersive}, and (ii)~the interaction of a trial soliton with a monochromatic KPII soliton gas, obtained as the $M=1$ reduction of the polychromatic ansatz. 
In both cases, the kinetic predictions agree with the numerical results, providing strong validation of the theory.
%

The setup (i) is related to the study of the oblique interactions between solitons and mean flow in \cite{NLTY2021v34p3583}.
Note, however, that in \cite{NLTY2021v34p3583} the initial condition was a half soliton, initially placed to the left or to the right of the step that gives rise to the rarefaction wave (RW) and DSW, allowing for the possibility of the soliton being trapped inside the RW or DSW during evolution.
Here, in contrast, we are considering exact multi-soliton solutions of the KPII equation, so the trial soliton always reappears on the other side of the condensate.  

The present results suggest several directions for future work. 
First, the present theory describes \emph{non-resonant} gases, in which every pair of line solitons violates condition~\eqref{eq:rescond}.
A kinetic description that accommodates a finite density of Y-shaped Miles resonances~\cite{miles1977diffraction,biondini2003family,kodama2004young,kodama2017kp} and of the associated web structures would describe a richer class of 2D nonlinear wave fields.
Second, as in the 1D setting~\cite{bonnemain2022generalized,doyon_generalized_2025}, the present formulation should provide a starting point for a full thermodynamics of 2D soliton gases,
now over an effective 2D spectral space parameterized jointly by amplitude~$a$ and slope~$c$. 
Third, since the geometric argument underlying the equality of the $t_2$- and $t_3$-shifts is generic across integrable hierarchies, an analogous theory should exist for other (2+1)-dimensional integrable dispersive models such as the Davey--Stewartson equation.
Fourth, a rigorous mathematical justification of the kinetic equations as a large-$N$ limit of exact $N$-soliton  solutions of the KPII equation is an obvious challenge -- we believe that the techniques of \cite{doyon2026solitonskdvsolitongas}, which give unambiguous positions of KdV solitons everywhere in space-time, can be generalised to the line solitons of KPII, because the KPII tau function takes a similar form as that of KdV. Fifth, it would be natural to extend the known results on diffusive corrections, correlation functions and large deviations from 1D integrability, to the setup of 2D line soliton gases. Finally, the recent experimental realization of random 2D shallow-water soliton gases~\cite{leduque2024space} suggests that it may be possible to experimentally test some of the predictions of the present theory, most directly, the effective-slope and effective-velocity renormalization in the monochromatic-gas reduction.

\textit{Acknowledgments.}
GB\ was partially supported by the Simons Foundation under grant number SFI-MPS-TSM-00013369. BD was supported by the EPSRC grant EP/Z534304/1, under the UK Research and Innovation Horizon Europe Guarantee, Advanced Grant Scheme. GE was supported by the EPSRC grant EP/C002401/1.


\bibliographystyle{apsrev4-1}
\bibliography{Biblio,Biblio-2019-01-03,Bibliography}

\section*{End Matter}

\paragraph{Multiple-time compatibility.}
Equations \eqref{ceff} are the ``effectivisation'' of the velocity $v_1(a,c) = c$ and $v_2(a,c) = 4a^2 + c^2$, respectively. One of the basic results of GHD \cite{doyon2020lecture} is that one can express such effective velocities in terms of the inverse of an integral operator, acting on functions of $a,c$, constructed from the scattering shift function. That is,
\begin{equation}
    v_i^{\rm eff}(a,c) =
    \frac{(\boldsymbol 1-\mathcal T n)^{-1}v_i(a,c)}{1 + \int \dd a'\dd c'\,
    \rho_{a',c'}\varphi(a,c;a',c')}
\end{equation}
where $\mathcal T n$ is the integral operator
\begin{equation}
    \mathcal T n f(a,c) =
    \int \dd a'\dd c'\,
    \varphi(a,c;a',c')
    n_{a',c'}
    f(a',c')
\end{equation}
where
\begin{equation}
    n_{a,c}=
    \frac{\rho_{a,c}}{
    1 + \int \dd a'\dd c'\,
    \rho_{a',c'}\varphi(a,c;a',c')
    }.
\end{equation}
is the so-called ``occupation function''.
Therefore, the compatibility condition \eqref{consist} is
\begin{equation}\label{consistem}
    \partial_{t_1} \Big(n_{a,c}(\boldsymbol 1-\mathcal T n)^{-1}v_2(a,c)\Big)
    =
    \partial_{t_2}\Big(
    n_{a,c}(\boldsymbol 1-\mathcal T n)^{-1}v_1(a,c)\Big).
\end{equation}
Note that the operator $\mathcal Tn$ clearly has the form of an integral operator $\mathcal T$ times the multiplication operator, which multiplies point-wise by $n_{a,c}$. 

Another fundamental result \cite{doyon2020lecture} is that the continuity equations \eqref{conty} are equivalent to the following transport equations for $n_{a,c}$:
\begin{equation}
    \partial_{t_i} n_{a,c} + v_i^{\rm eff}(a,c)\partial_x n_{a,c} = 0.
\end{equation}
With this, and using the operator relation $\partial_{t_i} \big(n (\boldsymbol 1-\mathcal T n)^{-1}\big) = (\boldsymbol 1- n\mathcal T)^{-1} \partial_{t_i} n (\boldsymbol 1-\mathcal T n)^{-1}$, we have
\begin{eqnarray}
    \lefteqn{
    \partial_{t_i} \Big(n(\boldsymbol 1-\mathcal T n)^{-1}v_j\Big)}
    &&\\
    &=&
    (\boldsymbol 1- n\mathcal T)^{-1}\partial_x n\,\Big(
    \big((\boldsymbol 1-\mathcal T n)^{-1}v_i\big)\big((\boldsymbol 1-\mathcal T n)^{-1}v_j\big)
    \Big).\nonumber
\end{eqnarray}
As the right-hand side is symmetric, it follows that the compatibility condition \eqref{consistem} is satisfied.

\medskip
\paragraph{Alternate derivation of the kinetic equations.}
Another way to derive the kinetic equations \eqref{conty} and~\eqref{consist} is a generalization of the original 1D derivation for the KdV equation~\cite{el2003thermodynamic, el2021soliton}. It follows from the consideration of the thermodynamic limit of the slow modulation (Whitham) equations for any integrable 2D dispersive hydrodynamics admitting multiphase wave solutions of the form
\vspace*{-0.6ex}
\begin{equation}\label{multiph}
u(x,y,t) =\Phi(\theta_1, \theta_2, \dots, \theta_N), \ \  \theta_j = k_jx +l_j y - \omega_j t,
\end{equation}
where the parameters $k_j, l_j$, and $\omega_j$,  $j=1, \dots, N$ are respectively the components of the 2D wave vector and frequency associated with each phase. 
When slow $xyt$-modulations of the wave parameters in \eqref{multiph} are allowed,
one can define the local wavevector and local frequencies by \cite{whitham2011linear}
\vspace*{-0.6ex}
\begin{equation}\label{local}
k_j :=\theta_{j, x}, \quad l_j :=\theta_{j,y}, \quad \omega_j :=-\theta_{j,t} \, .
\end{equation}
The consistency of partial derivatives in \eqref{local} then immediately implies the modulation equations
\begin{equation}\label{mod}
k_{j,t} + \omega_{j,x}=0, \quad  \quad k_{j,y} - l_{j,x} =0\, , \quad l_{j,t} + \omega_{j,y}=0\, ,
\end{equation}
Equations \eqref{mod} are part of the full system of modulation equations for the integrable 2D equation considered. 
Indeed, Eqs.~\eqref{mod} with $N=1$ were used to derive the one-phase Whitham modulation equations for the KP equation in \cite{ablowitz2017whitham}.
The full system of modulation equations also includes equations for slow evolution of other parameters of the multiphase solution \eqref{multiph}, such as the amplitudes of the nonlinear wavemodes, the wave mean, etc. 

The $N$-soliton limit of the multiphase solution \eqref{multiph} is realized by letting
$k_j \to 0$, $l_j \to 0$ and $\omega_j \to 0$, for $j=1, \dots, N$. 
In this limit, the total density of solitons $\sum_{j=1}^N k_j/(2\pi) \to 0$.  
In contrast, to obtain a thermodynamic-type {\it soliton gas limit}, 
we consider the limit $N\to\infty$ with 
\vspace*{-0.6ex}
\be \label{therm}
k_j \sim l_j \sim \omega_j  \sim \frac{1}{N},
\ee
so that the total density of solitons  remains finite. 
Next, parameterizing the solitons by their amplitudes $a$ and slopes $c$, we introduce the density of states $\rho_{a,c}$  and the $x$- and $y$- flux densities $g_{a,c}= \veff(a,c)\rho_{a,c}$ and $q_{a,c}=\ceff(a,c)\rho_{a,c}$ respectively via 
\vspace*{-0.5ex}
\bse
\label{therm1}
\begin{gather}
\frac{1}{2\pi}\sum_{j=1}^M k_j \to \int^c \int^a \rho_{a'c'} \dd a'\dd c', \\
\frac{1}{2\pi}\sum_{j=1}^M \omega_j \to \int^c \int^a g_{a'c'} \dd a'\dd c', \\ 
\frac{1}{2\pi}\sum_{j=1}^M l_j \to \int^c \int^a q_{a'c'} \dd a'\dd c'\, ,
\end{gather}
\ese
where $M\le N$, and where we assumed the thermodynamic scaling \eqref{therm}. Note the definition  of the density of states $\rho_{a,c}$ in  \eqref{therm1}a is consistent with the phenomenological definition of the density of line solitons in \eqref{dos}.
Introducing a slow spatiotemporal dependence $\rho_{a,c}(x,y,t)$, $g_{a,c}(x,y,t)$, $q_{a,c}(x,y,t)$ and applying the thermodynamic limit specified by \eqref{therm} and \eqref{therm1} to the modulation system \eqref{mod} we obtain the kinetic equations  \eqref{conty} and the consistency condition \eqref{consist}. The equations of state \eqref{ceff} for $\ceff(a,c)$, $\veff(a,c)$ providing a closure for the kinetic system \eqref{conty}, \eqref{consist} can be introduced phenomenologically, via a collision rate ansatz,  as it is done in the main text. Their systematic derivation 
involves a rather complicated system-specific analysis of nonlinear dispersion relations, see \cite{el2003thermodynamic,el2020spectral, el2021soliton}. We leave the generalization of this derivation for the KPII equation SG to future works.

\medskip
\paragraph{Polychromatic reductions.}
Under the polychromatic ansatz $\rho_{a,c}(x,y,t) = \sum_{j=1}^M w_j(x,y,t)\delta(a-a_j)\delta(c-c_j)$ (see \cite{el2011kinetic, bonnemain2025two}), the kinetic equations \eqref{conty}, \eqref{consist} reduce to $M$ coupled two-dimensional systems of hydrodynamic type 
\vspace*{-1ex}
\bse
\begin{gather}
    \partial_y w_j + \partial_x\left(\ceff_j w_j\right) = 0 \\
    \partial_t w_j + \partial_x\left(\veff_j w_j\right) =0 \\
    \partial_t \left(\ceff_j w_j\right) + \partial_y\left(\veff_j w_j\right) =0
\end{gather}
\ese
for $j=1, \dots, M$,
with the reduced effective slopes $\ceff_j \equiv \ceff(a_j,c_j)$ given by
\vspace*{-0.6ex}
\begin{equation}
    \ceff_j = \frac{c_j + \sum_{k\neq j} w_k \varphi_{jk} \ceff_k}{1 + \sum_{k\neq j} w_k \varphi_{jk}} \, , 
\end{equation}
and effective velocities $\veff_j \equiv \veff(a_j,c_j)$ given by
\vspace*{-0.6ex}
\begin{equation}
    \veff_j = \frac{v_j +  \sum_{k\neq j} w_k \varphi_{jk} \veff_k}{1 + \sum_{k\neq j} w_k\varphi_{jk}} \, , 
\end{equation}
where $\varphi_{jk} \equiv \varphi(a_j,c_j;a_k,c_k)$. In particular, when considering the interaction of a single trial soliton  with a monochromatic gas with soliton parameters $(a_1, c_1)$ and density $w_1$, one should take $M=2$ and the limit $w_2 \to 0$.
Then, one obtains
\bse \label{bichrom_eff}
\begin{gather}
    \ceff_1 = c_1 \, ,\qquad \veff_1 = v_1 \, ,
    \\[0.2ex]
    \ceff_2 = \frac{c_2 + w_1 \varphi_{21} c_1 }{1+w_1\varphi_{21}}\, ,\qquad \veff_2 = \frac{c_2 + w_1 \varphi_{21} v_1}{1+w_1\varphi_{21}}\, ,
\end{gather}
\ese
where $(a_2, c_2)$ are the parameters of the trial soliton.

\end{document}


\pagestyle{fancy}
\renewcommand{\headrulewidth}{0pt}
\fancyhead{ }
\fancyfoot[c]{\small\thepage/\pageref*{LastPage}}
\maketitle

\section{Interaction-induced position shifts in the Kadomtsev-Petviashvili II equation revisited}

In this section notes is to review the calculation of the position shift of an asymptotic line soliton of the Kadomtsev-Petviashvili~II (KPII) equation 
resulting from interaction with another line soliton after evolution with respect to either the second spatial variable~$y$ or the temporal variable~$t$, following~\cite{biondini2007line}.

\subsection{Preliminaries: Multi-soliton solutions of KPII}

\textbf{KPII, tau function and Wronskian solutions.}
For the present purposes, it is convenient to write the KPII equation as
\vspace*{-0.4ex}
\[
( u_t + 6 u u_x + u_{xxx} )_x +  u_{yy} = 0\,,
\label{e:KP}
\]
so we can use the formalism of \cite{biondini2006soliton} without change.
Apart from the inessential numerical constants, the main change from the ``physical'' normalization of the KP equation is that the coefficient of the temporal derivative is negative, so the temporal variable in~\eqref{e:KP} flows in the opposite direction as the physical one.) 

Multi-soliton solutions of~\eqref{e:KP} can be obtained via \cite{biondini2006soliton}
\vspace*{-0.8ex}
\[
u(x,y,t) = 2\partialderiv[2]{ }x \log\tau_{N,M}(x,y,t)\,,
\label{e:wrt}
\]
with
\vspace*{-0.8ex}
\bse
\label{e:tau}
\begin{gather}
\tau_{N,M}(x,y,t) = \wr(f_1,\dots,f_N)\,, 
\\
( f_1,\dots,f_N )^T = G\,( \e^{\theta_1},\dots,\e^{\theta_M} ) ^T\,,
\\
\theta_m = k_m x + \sqrt{3}k_m^2 y -4 k_m^3 t\,,
\end{gather}
\ese
and where without loss of generality one can take the ``phase parameters'' $k_1,\dots,k_M$ to be well-ordered: $k_1 < k_2 < \cdots < k_M$.
The $N\times M$ matrix $G = ( g_{nm} )$ is referred to as the coefficient matrix.
The solution is non-singular as long as all principal minors of $G$ are non-negative.

\textbf{Single line soliton, soliton parameters.}
The simplest nontrivial case $N = 1$ and $M = 2$ generates the 1-soliton solution $u(x,y,t) = u_{12}(x,y,t)$, with
\vspace*{-0.4ex}
\[
u_{ij}(x,y,t) = 2 a_{ij}^2 \sech^2[\Delta_{ij}]\,, 
\qquad 
\Delta_{ij} = \half(\theta_i - \theta_j) =  a_{ij}(x - q_{ij} y - v_{ij} t)\,,
\label{e:KPsoliton}
\]
where
\vspace*{-0.4ex}
\begin{gather}
\qquad
a_{ij} = \half \left(k_j - k_i\right),
\qquad
q_{ij} = -\sqrt{3}\left(k_i+ k_j\right),
\qquad
v_{ij} = 4\left(k_i^2 + k_ik_j + k_j^2\right)\,.
\label{e:solitonparameters}
\end{gather}
We call $a_{ij} $ and $q_{ij}$ the amplitude and slope parameters, respectively, and $v_{ij}$ the soliton speed.
(Again, note that the direction of the time variable in~\eqref{e:KP} is reversed compared to the physical coordinate.)
Note also that $q_{ij} = \tan\varphi_{ij}$ where $\varphi_{ij}$ is the angle that the soliton makes with the positive $y$-axis, counted counterclockwise.

\textbf{Multi-soliton solutions, dominant phase combinations, index pairs.}
For $N>1$, the Binet-Cauchy theorem allows one to show that
\[
\tau_{N,M}(x,y,t) = \det( K \e^\Theta G) =
\sum_{1\le m_1 < m_2 < \cdots < m_N \le M} d_{m_1,\dots,m_N}G_{m_1,\dots,m_N}\e^{\theta_{m_1,\dots,m_N}}\,,
\label{e:tauNM}
\]
where 
$K = ( k_m^{n-1} )$, 
$\Theta = (\theta_1,\dots,\theta_M)$,
$G_{m_1,\dots,m_N}$ is the $N\times N$ principal minor obtained by selecting columns $m_1,\dots,m_N$ of~$C$, 
$d_{m_1,\dots,m_N}$ is a Van der Monde determinant, 
and 
$\theta_{m_1,\dots,m_N} = \theta_{m_1} + \cdots + \theta_{m_N}\,$.
That is, the tau function is a sum of exponential phase combinations.
This observation is the key for the analysis that follows.

We call a phase combination $\theta_*$ \textit{dominant} in a region $R\in\Real^3$ if, $\forall (x,y,t)\in R$, $\theta_*$ is larger than any other phase combination appearing in the tau function.
It is easy to see that the corresponding solution of KP is exponentially small in any region where a single phase combination is dominant.
That is, the solution is localized along those lines where a balance exists between dominant phase combinations.
In particular, it was shown in \cite{biondini2006soliton} that, generically, dominant phase combinations in adjacent spatio-temporal regions have $N-1$ common phases.
That is, solitons are localized along lines $L_{ij}:\theta_i=\theta_j$ for suitable choices of indices $i\&j$.
Moreover, it was also shown in \cite{biondini2006soliton} that, up to reducible cases, \eqref{e:tau} generates a solution with $N$ asymptotic line solitons as $y\to\infty$ and $M-N$ asymptotic line solitons as $y\to-\infty$.
Each asymptotic line soliton has the form \eqref{e:KPsoliton}, with $i,j$ two distinct indices between 1 and~$M$.
Thus, each soliton can be uniquely identified by an \textit{index pair} $[i,j]$.

\textbf{Elastic two-soliton solutions.}
The case $M = 2N$ yields the $N$-soliton solutions of KPII.
Those $N$-soliton solutions for which the amplitudes and slopes of the asymptotic solitons as $y\to\infty$ are the same as those of the asymptotic solitons as $y\to-\infty$ are called \textit{elastic}.
It was shown in \cite{kodama2004young} that 3 inequivalent classes of elastic 2-soliton solutions of KPII exist
(labeled in \cite{biondini2007line} ``ordinary'', ``asymmetric'' and ``resonant''), corresponding to the 
coefficient matrices
\[
G_\ord = \begin{pmatrix} 1 & 1 & 0 & 0 \\ 0 & 0 & 1 & 1 \end{pmatrix},\qquad
G_\asymm = \begin{pmatrix} 1 & 0 & 0 & -1 \\ 0 & 1 & 1 & 0 \end{pmatrix},\qquad
G_\res = \begin{pmatrix} 1 & 0 & c_{23} & c_{24} \\ 0 & 1 & -1 & -1 \end{pmatrix}.
\]
The corresponding tau functions are:
\bse
\begin{gather}
\tau_\ord = d_{13} \e^{\theta_{13}} + d_{14} \e^{\theta_{14}} + d_{23} \e^{\theta_{23}} + d_{24} \e^{\theta_{24}}\,,
\label{e:tauord}
\\
\tau_\asymm = d_{12} \e^{\theta_{12}} + d_{13} \e^{\theta_{13}} + d_{24} \e^{\theta_{24}} + d_{34} \e^{\theta_{34}}\,,
\\
\tau_\res = d_{12} \e^{\theta_{12}} + d_{13} \e^{\theta_{13}} + d_{14} \e^{\theta_{14}} + d_{23} \e^{\theta_{23}} + d_{24} \e^{\theta_{24}} + d_{34} \e^{\theta_{34}}\,,
\end{gather}
\ese
where now $d_{ij} = k_j - k_i$.
The corresponding asymptotic line solitons are identified by the index pairs:
\[
\ord:~[1,2]~\&~[3,4]\,;\qquad
\asymm:~[1,4]~\&~[2,3]\,;\qquad
\res:~[1,3]~\&~[2,4]\,.
\]
It was then shown in \cite{biondini2007line} that these three classes of solutions cover disjoint sectors of the physical 2-soliton parameter space.

\subsection{Spatial evolution}

To compute the interaction-induced position shift, one must look in detail at which phase combinations are dominant in different regions of the $xy$-plane.

We consider the asymptotics as $y\to\pm\infty$ along the lines $L_q:x = \xi - q y$ with $\xi = O(1)$.
Similarly to before, $q = \tan\varphi$, where $\varphi$ is the angle that $L_q$ makes with the positive $y$ axis, counted counterclockwise.
Moreover,
$\theta_i = k_i\xi +\sqrt{3} k_i (k_i - q) y -4 k_i^3 t$,
implying
\begin{gather}
\Delta_{ij} = \half(k_i - k_j) \big[ \xi + \sqrt{3}( k_i + k_j - q ) y -4 (k_i^2 + k_i k_j + k_j^2) t \big]\,.
\label{e:thetaijy}
\end{gather}
One can now use~\eqref{e:thetaijy} to:
(i) identify the dominant phase combinations as $y\to\pm\infty$ as a function of $q$,
(ii) identify the transitions between these dominant phase combinations, thereby recovering the asymptotic line solitons, and 
(iii) obtain the interaction-induced position shift.
We show how this works and do the calculations explicitly for the ordinary soliton solutions.
The calculations for the other 2 classes of solutions are entirely analogous.

\begin{figure}[b!]
\centerline{%
\includegraphics[width=0.325\textwidth]{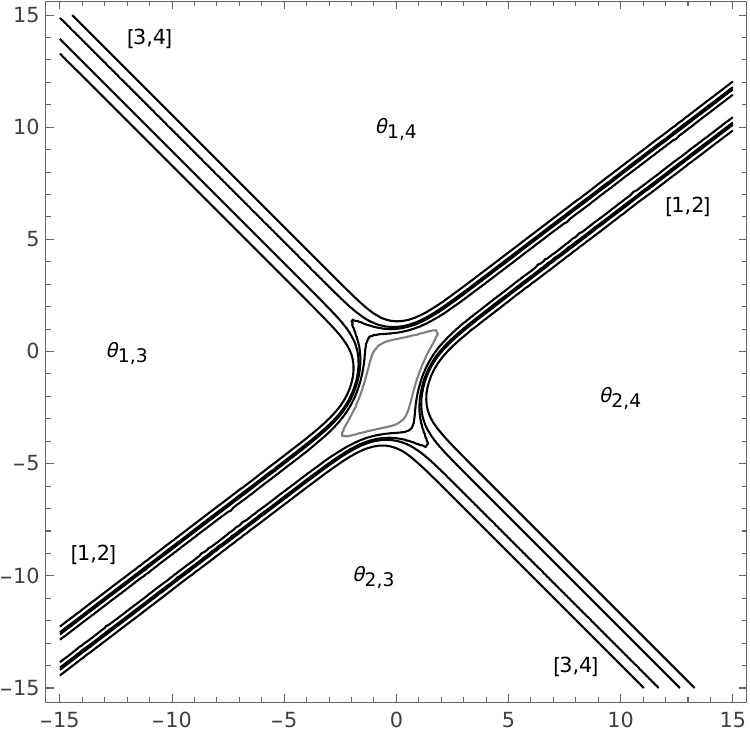}\qquad
\includegraphics[width=0.325\textwidth]{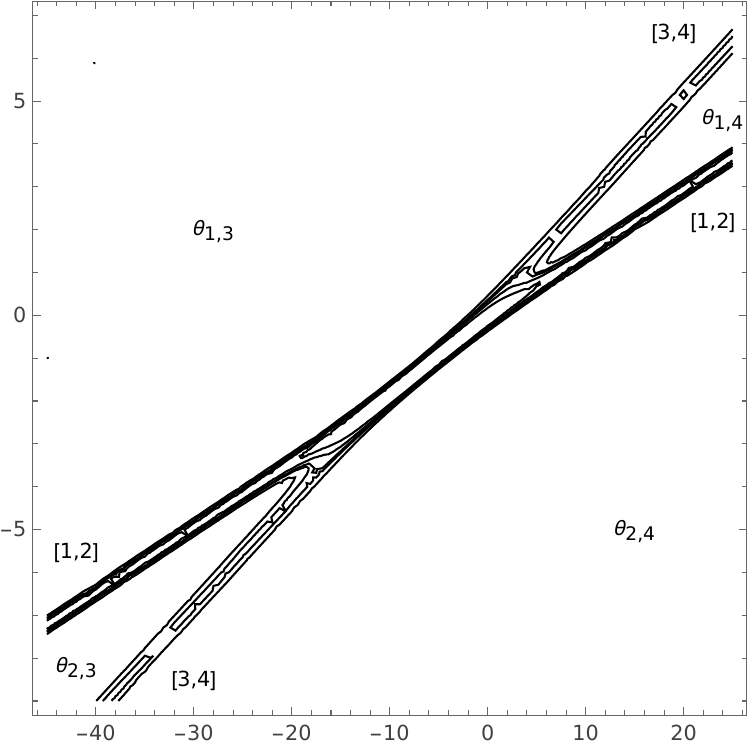}
}
\caption{
Left: The asymptotic line solitons and the dominant phase combinations in the $xy$-plane for an ordinary 2-soliton solution.
Right: Same, but in the $xt$-plane.
(Note that, since ordinary 2-soliton solutions are traveling wave solutions, time evolution in the left plot and $y$ evolution in the right plot amount to just an overall translation of the solution in each plot, cf.\ Fig~\ref{f:2} below.)}
\label{f:1}
\kern-\medskipamount
\end{figure}

\textbf{Asymptotics as $y\to\infty$.}
Recall that the tau function for ordinary 2-soliton solutions is given by~\eqref{e:tauord}.
As $y\to\infty$, $\theta_{13}$ is the dominant phase combination in $\tau_\ord$ for sufficiently large positive values of $q$.
(Cf.\ Fig.~1a in \cite{biondini2006soliton} and Fig.~\ref{f:1}a below.)
More precisely, $\theta_{13}$ is the dominant phase combination for all $q > q_{34}$.  
This is because, for $q = q_{34}$, one has $\theta_3 - \theta_4 = O(1)$, which in turn implies 
$\theta_{1,3} - \theta_{1,4} + O(1)$, so there is a dominant balance.
For $q < q_{34}$, $\theta_{14}$ is dominant.
This situation continues until $q = q_{12}$, at which point $\theta_1 - \theta_2 = O(1)$ resulting in another dominant balance.
Finally, for $q < q_{12}$, $\theta_{24}$ is dominant.
Hence, as $y\to\infty$, the asymptotic solitons are localized along the lines $L_{12}\&L_{34}$, and are therefore identified by the index pairs $[1,2]\&[3,4]$ (as anticipated above).

As $y\to\infty$ in a neighborood of $L_{12}$, we have, up to exponentially smaller terms,
\[
\tau = d_{14} \e^{\theta_{14}} + d_{24} \e^{\theta_{24}}
  = \e^{\theta_4 + \frac12\theta_{12}} \big( d_{14}d_{24} \big)^{1/2} 2\cosh\big(\Delta_{12} + \delta_{12}^+ \big)\,,
\]
where $\delta_{12}^+ = \half\log(d_{14}/d_{24})$.
The corresponding asymptotic soliton is (again up to exponentially smaller terms)
\[
u_{12}^+ = 2 a_{12}^2 \sech^2 \big[\Delta_{12} + \delta_{12}^+ \big]\,. 
\label{e:u12+}
\]
\textbf{Asymptotics as $y\to-\infty$.}
Next we perform similar calculation as $y\to-\infty$ in order to obtain the position shift.
As $y\to\infty$, $\theta_{13}$ is the dominant phase combination in for all $q > q_{12}$.
For $q = q_{12}$ one has $\theta_2 - \theta_1 = O(1)$, so there's again a dominant balance,
and for $q < q_{12}$ the dominant phase combination is $\theta_{23}$ until $q = q_{34}$.
Thus, as $y\to-\infty$ in a neighborhood of $L_{12}$ we have, up to exponentially smaller terms,
\[
\tau = d_{13} \e^{\theta_{13}} + d_{23} \e^{\theta_{23}}
  = 2 \e^{\theta_3 + \frac12\theta_{12}} \big( d_{13}d_{23} \big)^{1/2} \cosh\big(\Delta_{12} + \delta_{12}^- \big)\,,
\]
where $\delta_{12}^- = \half\log(d_{13}/d_{23})$.
The corresponding asymptotic soliton is (again up to exponentially smaller terms)
\[
u_{12}^- = 2 a_{12}^2 \sech^2 \big[\Delta_{12} + \delta_{12}^- \big]\,. 
\]

\textbf{Position shift.}
The resulting position shift of soliton [1,2] from the interaction with soliton [3,4] is thus:
\[
 \delta x_{12} = \frac{\delta_{12}^+ - \delta_{12}^-}{\half(k_2-k_1)} = \frac1{k_2-k_1} \log \frac{d_{14}d_{23}}{d_{24}d_{13}}\,.
\]
Note
\[
\frac{d_{14}d_{23}}{d_{24}d_{13}} = \frac{(k_4-k_1)(k_3-k_2)}{(k_4-k_2)(k_3-k_1)} 
  = \frac{ ( a_{12} - a_{34} )^2 - ( q_{12} + q_{34} )^2 } { ( a_{12} - a_{34} )^2 - ( q_{12} - q_{34} )^2 }
\,,
\]
where we inverted \eqref{e:solitonparameters} to express $k_1,\dots,k_4$ in terms of $a_{12},a_{34},q_{12}\&q_{34}$.
Once the correct signs are identified, the above expression should coincide with equation~(8) in \cite{biondini2007line}.

\subsection{Temporal evolution}

We now perform a similar analysis in the $xt$-plane in order to compute the spatial position shift upon evolution along the temporal variable.
Namely, we consider the asymptotic behavior as $t\to\pm\infty$ along lines $L_v:x = \xi - v t$, again with $\xi = O(1)$.
Similarly to before, $v = \tan\alpha$, where $\alpha$ is the angle that $L_v$ makes with the positive $t$ axis, counted counterclockwise.
We now have
$\theta_i = k_i\xi +\sqrt{3} k_i^2 y -4 k_i(k_i^2 - v)t$,
which implies
\begin{gather}
\Delta_{ij} = \half(k_i - k_j) \big[ \xi + \sqrt{3}( k_i + k_j ) y - 4 \big( k_i^2 + k_i k_j + k_j^2 - v \big ) t \big] \,.
\label{e:thetaijt}
\end{gather}
Similarly to before, we now use~\eqref{e:thetaijt} to identify the dominant phase combinations as $t\to\pm\infty$ as a function of~$v$,
again 

As $t\to\infty$, we have that $\theta_{13}$ is the dominant combination for all sufficiently large positive values of~$v$.
(Cf.\ Fig.~\ref{f:1}b above.)
More precisely, $\theta_{13}$ is dominant until $v = v_{12}$, at which point $\theta_2 - \theta_1 = O(1)$, implying $\theta_{13} - \theta_{23} = O(1)$, leading to a dominant balance.
For $v_{12} < v < v_{34}$, $\theta_{23}$ is dominant.
At $v = v_{34}$,  $\theta_3 - \theta_4 = O(1)$, implying $\theta_{23} - \theta_{24} = O(1)$, leading again to a dominant balance.
Finally, for $v > v_{34}$, $\theta_{34}$ is dominant.

As $t\to\infty$ along $L_{v_{12}}$ we have, up to exponentially smaller terms,
\[
\tau =  d_{13} \e^{\theta_{13}} + d_{23} \e^{\theta_{23}}
  = \e^{\theta_3 + \frac12\theta_{12}} \big( d_{13}d_{23} \big)^{1/2} 2\cosh\big(\Delta_{12} + \delta_{12}^+ \big)\,,
\]
implying that the solution is again given by~\eqref{e:u12+}, 
but now with $\delta_{12}^+ = \half\log(d_{13}/d_{23})$. 
Note that, in the spatial analysis, the expression on the RHS of $\delta_{12}^+$ yielded $\delta_{12}^-$ instead.

Similarly, as $t\to-\infty$ along $L_{v_{12}}$ (cf.\ Fig.~\ref{f:1}b), we have, to leading order,
\[
\tau = d_{14} \e^{\theta_{14}} + d_{24} \e^{\theta_{24}}
  = \e^{\theta_4 + \frac12\theta_{12}} \big( d_{14}d_{24} \big)^{1/2} 2\cosh\big(\Delta_{12} + \delta_{12}^- \big)\,,
\]
where $\delta_{12}^- = \half\log(d_{14}/d_{24})$, and again note that the expression on the RHS yielded $\delta_{12}^+$ before.

Therefore, the position shifts arising from evolution with respect to $y$ and $t$ are equal and opposite.

\begin{figure}[t!]
\centerline{\includegraphics[width=0.85\textwidth]{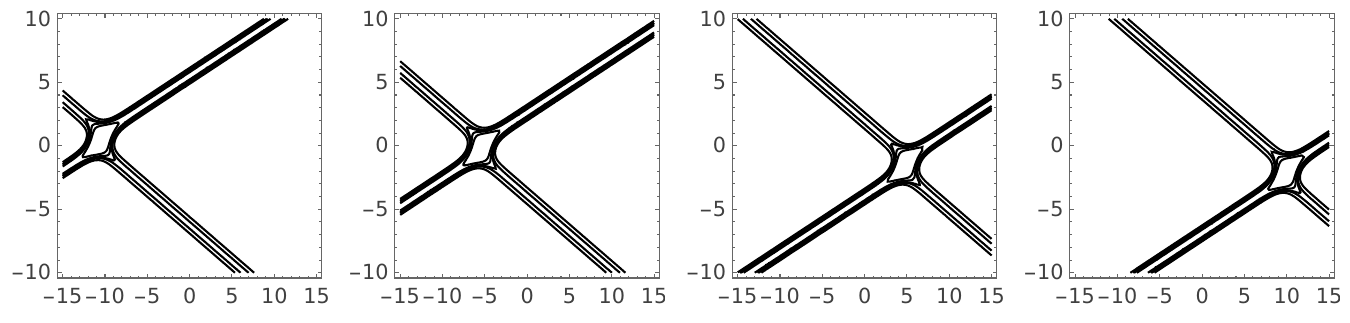}}
\centerline{\includegraphics[width=0.85\textwidth]{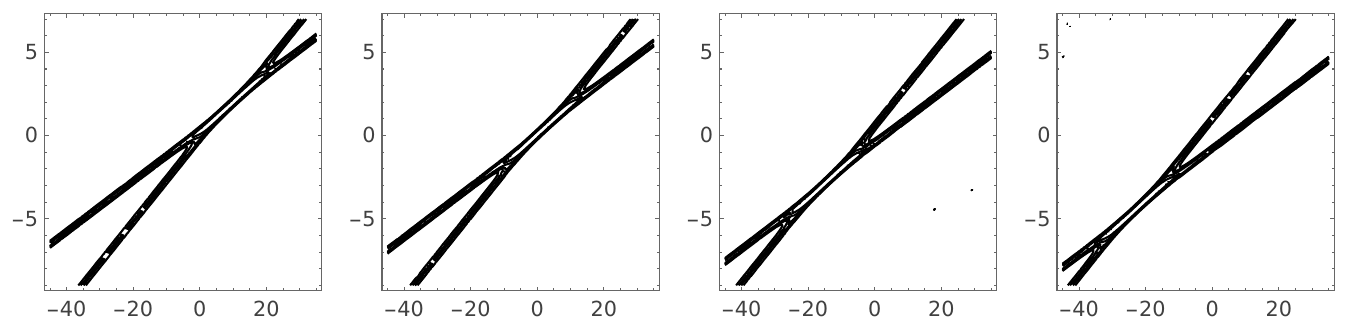}}
\caption{
Top row: The ordinary 2-soliton solution in Fig.~\ref{f:1}a in the $xy$-plane at $t=-2$, $-1$, 1 and~2.
Bottom row: Same, but for Fig.~\ref{f:1}b in the $xt$-plane for $y=-2$, $-1$, 1 and~2.
}
\label{f:2}
\end{figure}

\section{Detailed calculation of effective slopes}

We now discuss how to compute the effective slope in Figure 2 of the main text. At $t=0$ we consider a trial soliton of amplitude $a_{*}$ and slope $c_{*}$ interacting with a  1D dense SG exclusively  with slope $c_0 = 0$. The SG is initially characterized by the following (Weyl's) density of states
\begin{equation}
\label{eq:weyl}
    \rho(a) = \frac{a}{\pi\sqrt{1-a^2}} \; , \quad\quad a \in [0,1] \; , 
\end{equation}
which corresponds to a genus zero soliton condensate of the KdV equation with amplitude $1$ \cite{congy2023dispersive}. Note that interactions between solitons forming this gas do not affect their effective slope, indeed, $\ceff(a,0) = 0$ is solution of
\begin{equation}
    \ceff(a,c_0) = c_0 + \int_0^1 \dd a' \, \rho(a')\varphi(a,a';c_0,c_0)\left[\ceff(a',c_0)-\ceff(a,c_0)\right] \; ,
\end{equation}
where the interaction kernel reduces to the KdV one:
\begin{equation}\label{shift}
    \varphi(a,a';c_0,c_0) = \frac{1}{2a}\log\left|\frac{(c_0-c_0)^2 - 12 (a-a')^2}{(c_0-c_0)^2 - 12 (a+a')^2}\right| =  \frac{1}{a}\log\left|\frac{a-a'}{a+a'}\right| \; .
\end{equation}
Showing that $\ceff(a,0) = 0$ is actually the only solution of the above equation requires the introduction of an object analogous to the nonlinear dispersion relations of soliton gases \cite{el2021soliton}, which we leave for future publication.

We now present the procedure to evaluate the effective slope of the trial soliton at $t=0$. The equation of state (4) in the main text yields
\begin{equation}\label{Ceff}
    \ceff(a_{*},c_{*}) = c_{*}+ \frac{1}{2a_{*}}\int_0^1\dd a'\, \rho(a')\varphi(a_{*},a';c_{*},c_0)\left[\ceff(a_{*},c_{*})-c_0\right] \; ,
\end{equation}
from which we deduce
\begin{equation}\label{CeffInit}
    \ceff(a_{*},c_{*}) = \frac{c_{*}}{1 - \int_0^1\dd a'\, \frac{a'}{\pi\sqrt{1-a'^2}}\log\left(\frac{c_{*}^2-12(a_{*}-a')^2}{c_{*}^2-12(a_{*}+a')^2}\right) }\; ,
\end{equation}
where we assumed, for the sake of simplicity, that $a_{*}> 1$ and $c_{*}^2 < 12(a_{*}-1)^2$ so that we do not need the absolute value in Eq.~\eqref{shift} to keep the argument of the logarithm positive. As such, to compute the effective slope, rather than solving an integral equation as Eq.~\eqref{Ceff} might suggest, all we need to do is evaluate the integral through integration by part and partial fraction decomposition. Let $\alpha_{*}= \sqrt{12} a_{*}$. Then
\begin{equation}\label{Integral1}
\begin{aligned}
    \int_0^1 \dd a' \frac{a'}{\pi\sqrt{1-a'^2}}\log&\left(\frac{c_{*}^2-12(a_{*}-a')^2}{c_{*}^2-12(a_{*}+a')^2}\right)    = \int_0^{\sqrt{12}} \dd \alpha \frac{\sqrt{1-\alpha^2}}{144\pi}\partial_\alpha\log\left(\frac{c_{*}^2-(\alpha_{*}-\alpha)^2}{c_{*}^2-(\alpha_{*}+\alpha)^2}\right)\\ 
    &=\int_0^{\sqrt{12}} \dd \alpha \frac{\sqrt{1-\alpha^2/12}}{\sqrt{12}\pi}\left(\frac{1}{\alpha-\alpha_{*}-c_{*}} - \frac{1}{\alpha+\alpha_{*}-c_{*}} +\frac{1}{\alpha-\alpha_{*}+c_{*}} -\frac{1}{\alpha+\alpha_{*}+c_{*}} \right) \, ,
\end{aligned}
\end{equation}
which is evaluated using the following identity
\begin{multline}
    \int \dd x \frac{\sqrt{1-x^2}}{x+y+z} = 
    \\
    -1 +\frac\pi2 \left(y + z -\sqrt{(1+y+z)(y+z-1)}\right)+\sqrt{(1+y+z)(y+z-1)}\,\mathop{\rm arccot}(\sqrt{(1+y+z)(y+z-1)}) \, .
\end{multline}
For positive times, assuming that the trial soliton has a negligible impact on the dynamics of the gas, we make use of results from \cite{congy2023dispersive} in which the authors showed that the Riemann problem for the KdV condensate is characterized by contact discontinuities where
\begin{equation}
    \rho(a,x,t) = \left\{\begin{aligned}
& 0 && \text{if }\quad x<0 && \text{(region I)} \\
& \frac{a}{\pi\sqrt{x/6t-a}} && \text{if }\quad 0<x<6t && \text{(region II)} \\
& \frac{a}{\pi\sqrt{1-a}} && \text{if }\quad 6t<x<L-6t && \text{(region III)} \\
& \rho_{g_1}(a,x,t) && \text{if }\quad L-6t<x<L+4t && \text{(region IV)} \\
& 0 && \text{if }\quad x>L+4t && \text{(region V)}
\end{aligned}\right.
\end{equation}
In regions I and V the slope of the trial soliton is trivially $c_{*}$, the soliton has either not yet interacted or finished interacting with the gas. In region III the effective slope is given by the previously computed value using Eq.~\eqref{CeffInit}, while in region II one needs to evaluate the same integral \eqref{Integral1} in which the upper bound has been replaced by $\sqrt{x/6t}$ according to the modulation dynamics highlighted in \cite{congy2023dispersive}. Remains then to consider region IV where the gas is locally described by a modulated genus 1 condensate or, equivalentlys, a dispersive shock wave. The corresponding DOS is given by 
\begin{equation}
    \rho_{g_1}(a;x,t) = \frac{ia\left[a^2-w^2(x,t)\right]}{\pi R(a;x,t)} \, ,
\end{equation}
where we introduced the three functions
\begin{equation}
\begin{aligned}
    &w^2(x,t) = \lambda_3^2(x,t)-(\lambda_3^2-\lambda_1^2)\frac{\text{E}(m(x,t))}{\text{K}(m(x,t)} \, , \\
    &m = \frac{\lambda_2^2(x,t)-\lambda_1^2(x,t)}{\lambda_3^2(x,t)-\lambda_1^2(x,t)} \, , \\
    &R(a;xt) = \sqrt{(a^2-\lambda_1^2(x,t))(a^2-\lambda_2^2(x,t))(a^2-\lambda_3^2(x,t))} \, ,
\end{aligned}
\end{equation}
defined in terms of the complete elliptic integrals $E$ and $K$ as well as in terms of three parameters $\{\lambda_j\}_{j=1}^{3}$ that solve the modulation equations
\begin{equation}
        \partial_t\lambda_j +V_j(\lambda_1,\lambda_2,\lambda_3)\partial_x\lambda_j =0\, ,
\end{equation}
with characteristic velocities 
\begin{equation}
    \left\{\begin{aligned}
        & V_1(\lambda_1,\lambda_2,\lambda_3) = 2(\lambda_1^2+\lambda_2^2+\lambda_3^2)+\frac{4(\lambda_2^2-\lambda_1^2)}{\text{E}(m)/\text{K}(m)-1}\, , \\
        & V_2(\lambda_1,\lambda_2,\lambda_3) = 2(\lambda_1^2+\lambda_2^2+\lambda_3^2)+\frac{4(\lambda_3^2-\lambda_2^2)(\lambda_2^2-\lambda_1^2)}{\lambda_3^2-\lambda_2^2-(\lambda_3^2-\lambda_1^2)\text{E}(m)/\text{K}(m)}\, , \\
        & V_3(\lambda_1,\lambda_2,\lambda_3) = 2(\lambda_1^2+\lambda_2^2+\lambda_3^2)+\frac{4(\lambda_3^2-\lambda_2^2)\text{K}(m)}{\text{E}(m)}\, .
    \end{aligned}\right.
\end{equation}
The effective slope is then obtained as
\begin{equation}\label{CeffInit}
    \ceff_{{\rm region IV}}(a_{*},c_{*}) = \frac{c_{*}}{1 - \int_{\mathcal A_{{\rm region IV}}}\dd a\, \rho_{g_1}(a;x,t)\log\left(\frac{c_{*}^2-12(a_{*}-a)^2}{c_{*}^2-12(a_{*}+a)^2}\right) }\; ,
\end{equation}
in which the integration domain is $\mathcal A_{{\rm region IV}} = [0,\lambda_1(x,t)]\cup[\lambda_2(x,t),\lambda_3(x,t)]$. The integral in the above expression is evaluated numerically.  Here $\lambda_1(x,t)=\lambda_1(x,0) = 0$, and $\lambda_3(x,t)=\lambda_3(x,0)=1$.

\section{Efficient numerical synthesis of a two-dimensional soliton gas}

The algorithm adopted to generate the $N$-soliton solutions, approximating the SG for $N\gg 1$, is analogous to the one implemented in \cite{bonnemain2025two}. It relies on the Wronskian formulation~\eqref{e:wrt}  where for $\tau(x,y,t) = \det(K\,\e^{\Theta}G^T)$. Note that the Binet-Cauchy formula, which depends on the computation of the Van der Monde determinants (\ref{e:tauNM}), is impractical as the number of non-zero minors increases exponentially with the number of solitons, making the algorithm unsuitable for the numerical study of solutions with $N\gg1$.

For the purposes of numerical generation, we find it convenient to 
label the phase parameters corresponding to the $n$-th soliton as
$k_{n,\pm}$ and to choose the coefficient matrix as,
so that the $(2n-1)$-th and $2n$-th column of $G$ correspond to the
$n$-th soliton.
Then, following the map derived in \cite{bonnemain2025two} between the multi-soliton solution of Boussinesq and the Galilean-boosted KPII equations, the matrix $G$ in \eqref{e:tauNM} can generically be put in the form
\begin{equation}
G=\begin{pmatrix}
g_{1,1} & g_{1,2} & 0 & 0 & \cdots & 0 & 0 \\  0 & 0 & g_{2,3} & g_{2,4} & \cdots & 0 & 0 \\ \vdots & \vdots & \vdots  &  \vdots & \ddots &  \vdots &  \vdots \\ 0 &0 & \cdots & \cdots & 0  &g_{N,2N-1} & g_{N,2N}
\end{pmatrix} \; ,
\end{equation}
where
\begin{equation}
g_{n,2n} = C_n \frac{g_{n,2n-1}}{k_{n,+}-k_{n,-}}\prod_{n\neq m}\frac{k_{n,-}-k_{m,-}}{k_{n,+}-k_{m,-}}
\end{equation}
with $C_n=k_{n,+}-k_{n,-} $. Without loss of generality, we can assume $g_{n,2n-1}=1$ for $n=1,2\ldots N-1$ and the element $g_{N,2N-1}$ is given by:
\vspace*{-1ex}
\begin{equation}
g_{N,2N-1}=2\prod_{n=N}^{1}\prod_{m=1}^{n-1} \sgn(k_{n,-}-k_{m,-}).
\end{equation}
Note that in the numerical implementation, we consider:
\be
\theta_m = k_m \left(x-x_{m,0}\right) + \sqrt{3} k_m^2 \left(y-y_{m,0}\right) - 4 k_m^3 t\,,
\ee
where $x_{m,0}$ is the positional phase parameter of the $m$-th soliton within the $N$-soliton solution at $y=y_{m,0}$ (for $N=1$ this gives the position of the maximum of the isolated line soliton at $y=y_{m,0}$).

The algorithm, as noted in \cite{bonnemain2025two}, is susceptible to round-off errors arising from the summation of exponentially small and large values. To address this, high-precision arithmetic routines have been implemented, following the procedure established in \cite{gelash2018strongly} for the numerical generation of $N$-soliton solutions for the NLS equation.

\section{Detailed parameters used to generate the SGs in the main text.}

\textit{Fig.~1.}
The parameters used for the numerical simulations shown in the figure are generated according to:
\begin{equation}
a_{2n}=\alpha a_{2(n-1)}, \qquad
a_{2n-1}=\alpha a_{2(n-1)-1}, \quad
k_{2n,+}=-k_{2n-1,-}=-\frac{n}{2}\,\delta k,
\end{equation}
with
$\delta k=10^{-12}$, and $n=1,\ldots,\frac{N}{2}$.

Top row: Deterministic SG realized via $N=40$-soliton solution with fixed parameters 
$a_1 =a_2 = 1$, $\alpha=0.9$ and $x_{n,0}=-128$, $y_{n,0}=0$.

Bottom row: Stochastic SG realized via $N=30$-soliton solution with random parameters given by
$a_1=1.05$, $a_2=1$, and $\alpha = 0.9 + r$, where $r$ is a uniformly distributed random variable in the interval
$r\in[-0.1,\,0.1]$. Furthermore, $x_{n,0} \in \left[-125;125\right]$ and $y_{n,0} \in \left[-1;1\right]$

\textit{Fig.~2.}
The parameters of the trial soliton are $a_{*}= 1.3001$ and $c_{*}=-1.03836$. The box is built as a genus zero soliton condensate  of the KdV equation, with amplitudes $a_{g,n}$ sampled from the Weyl’s density of states \eqref{eq:weyl}:
\begin{equation}
    a_{g,n}=\sqrt{1-\frac{n^2}{N^2}}.
\end{equation}

\textit{Fig.~3.}
The parameters of the trial soliton are $a_{*}= 1.3$, $c_{*}=0.1$. Following \cite{congy2024riemann}, the amplitude parameters of the solitons in the monochromatic gas are sampled from the DOS of  a genus 1 condensate of the KdV equation with spectral support $\Gamma = [0.885;0.915]$. The slopes of the solitons are chosen as $c_{g,n}=1.13+\delta_g$ with $\delta_g$ a uniformly distributed random variable in the interval $\delta_g\in [-5;5]\times10^{-5}$. The initial phases of solitons in the monochromatic gas are set as $x_{n,0}\in [-5; 5]$ and $y_{n,0}=0$.

\bibliographystyle{unsrt}
\bibliography{Biblio,Biblio-2019-01-03,Bibliography}